\documentclass{aa}
\usepackage{graphicx}
\usepackage{amssymb}
\usepackage{amsmath}
\begin{document}
\title{High accuracy transit photometry of the planet OGLE-TR-113b with
a new deconvolution-based method\thanks{Based on observations collected
with the SUSI2 imager at the NTT telescope (La Silla Observatory, ESO, Chile) in the programme 075.C-0462A.}}
\author{M.  Gillon  \inst{1, 2}  \and  F. Pont\inst{1}  \and   C. Moutou
\inst{3} \and   F. Bouchy\inst{4, 5} \and   F. Courbin\inst{6} \and   S. Sohy
\inst{2} \and P. Magain\inst{2}}
\offprints{M.  Gillon\\email : michael.gillon@obs.unige.ch}
\institute{Observatoire de Gen\`eve, 51 Chemin des Maillettes, CH-1290 Sauverny,
Switzerland \and Institut d'Astrophysique et de G\'eophysique,  Universit\'e
  de Li\`ege,  All\'ee du 6 Ao\^ut, 17,  Bat.  B5C, Li\`ege 1, Belgium \and
LAM, Traverse du Siphon, BP8, Les Trois Lucs, 13376 Marseille Cedex 12,
France \and Observatoire de Haute-Provence, 04870 St-Michel l'Observatoire, France 
\and Institut d'Astrophysique de Paris, 98bis Bd Arago, 75014 Paris, France
\and Laboratoire d'Astrophysique, Ecole Polytechnique F\'ed\'erale de Lausanne (EPFL), 
Observatoire, CH-1290 Sauverny, Switzerland}
\date{}
\abstract
    {A high accuracy photometry algorithm is needed to take full
advantage of the potential of the transit method for the
characterization of exoplanets, especially in deep crowded fields. It has to
reduce to the lowest possible level the negative influence of systematic
effects on the photometric accuracy.  It should also be able to cope with a
high level of crowding and with large scale variations of the spatial
resolution from one image to another. A recent deconvolution-based 
photometry algorithm fulfills all these requirements,
and it also increases the resolution of astronomical images, which is an
important advantage for the detection of blends and the discrimination of
false positives in transit photometry. We made some changes to this algorithm
in order to optimize it for transit photometry and used it to reduce NTT/SUSI2
observations of two transits of OGLE-TR-113b.  This reduction 
has led to two very high precision transit light curves with a low level of
systematic residuals, used together with former photometric and spectroscopic measurements to derive new stellar and planetary parameters in excellent agreement with previous ones, but significantly more precise.  

}
\authorrunning{M. Gillon et al.}
\titlerunning{High accuracy transit photometry of OGLE-TR-113b}
\maketitle
%
%
\section{Introduction}
Among the $\sim$200 exoplanets known so far, only the 10 ones transiting their
parent star have measured masses and radii, thanks to the complementarity of
the radial--velocity and transit methods. Among them, 5 were detected by the
OGLE-III planetary transit survey (\cite{Udalski1}, \cite{Udalski2},
\cite{Udalski3},  \cite{Udalski4}): OGLE-TR-10b (\cite{Konacki3}), OGLE-TR-56b
(\cite{Konacki1}, \cite{Bouchy2}), OGLE-TR-111b (\cite{Pont1}), OGLE-TR-113b 
(\cite{Bouchy1}, \cite{Konacki2}) and OGLE-TR-132b (\cite{Bouchy1}). Compared to 
the other
transiting exoplanets, they orbit much fainter stars, leading to a lower
amount of information available from their observation. Furthermore, obtaining
high accuracy photometry for these stars is difficult with a classical
reduction method, even with large telescopes, because of the high level of
crowding present in most of the deep fields of view in the Galactic plane.
Nevertheless, the accurate photometric monitoring of their transits is important to better constrain the mass-radius relationship of close-in giant planets, and thus the processes of planet formation, migration and evaporation. Besides, high accuracy
transit observations may allow the detection of other planets, even
terrestrial ones in the best cases, by the measurements of the dynamically
induced variations of the period of the transit (\cite{Miralda1},
\cite{Agol1}, \cite{Holman1}).

An image deconvolution algorithm (\cite{Magain1}) has recently been adapted to the photometric
analysis of crowded fields (\cite{Magain2}), even when the level of crowding is so high that no
isolated star can be used to obtain the $PSF$ (Point Spread Function).
We made some modifications to this algorithm to optimize it for follow-up transit photometry, with
a main goal in mind: to obtain the highest possible level of photometric accuracy, even for faint stars located in  deep crowded fields.

This new method was tested on new photometric observations of two OGLE-TR-113b
transits obtained with the NTT/SUSI2 instrument.  This planet was the second
one confirmed from the list of planetary candidates of the OGLE-III survey. It
orbits around a faint K dwarf star ($I$ = 14.42) in the constellation of
Carina. Due to the small radius of the parent star ($R \sim 0.8$ $R_{\odot}$),
the transit dip in the OGLE-III light curves is the largest one among the
planets detected by this survey ($\sim$ 3 \%). As OGLE-TR-113 lies in a field
of view with a high level of crowding, this case is ideal to validate the
potential of our new method. 

Sect.\  2 presents the observational data. Sect.\  3 summarizes the main
characteristics of the deconvolution algorithm and describes the improvements 
we brought to optimize it for follow-up transit photometry. In Sect.\  4, our
results are presented and new parameters are derived for the planet
OGLE-TR-113b. Finally, Sect.\  5 gives our conclusions.

\section{Observations}
The observations were obtained on April 3rd and 13th, 2005 with the SUSI2
camera on the ESO NTT (programme 075.C-0462A). In total, 235 exposures were
acquired during the first night, and 357 exposures during the second night, in
a 5.4$\arcmin$ $\times$ 5.4$\arcmin$ field of view. The exposure time was 32 s, the read-out time was 23 s, and the $R\#813$ filter was used for all observations. We used SUSI2 with a 2 $\times$ 2 pixel binning in order to get at the same time a good spatial and a good
temporal sampling. The binned pixel size is 0.16\arcsec. The measured seeing varies between 0.85$\arcsec$ and 1.38$\arcsec$ for the first night and between 1.29$\arcsec$ and 1.81$\arcsec$ for the second night. Transparency was high and stable for both nights. The airmass of the field decreases from 1.35 to 1.18 then grows to 1.20 during the first sequence, and decreases from 1.23 to 1.18 then grows to 1.54 during the second sequence. 

The frames were debiassed and flatfielded with the standard ESO pipeline. 

In addition to these new data, we used  VLT--FLAMES radial velocity
measurements  (\cite{Bouchy1}), stellar parameters derived from VLT-UVES
spectra (\cite{Santos1}) and OGLE-III ephemeris (\cite{Konacki2}) to constrain
the physical and orbital parameters of OGLE-TR-113b.
\section{Photometric reduction method}
\subsection{MCS deconvolution algorithm}
The MCS deconvolution algorithm (\cite{Magain1}, hereafter M1) is an
image processing  method specially adapted to astronomical images containing
 point sources, which allows to achieve (1) an increase of the angular resolution, (2) an
accurate determination of the positions (astrometry) and the intensities
(photometry) of the objects lying in the image. One of its main characteristics
is to perform a \emph{partial} deconvolution, in order to obtain a final image 
in agreement with the sampling theorem (\cite{Shannon1}, \cite{Press1}).  This
partial deconvolution is done by using, instead of the total $PSF$, 
a partial $PSF$ which is a convolution kernel connecting the deconvolved image to the
original one.

In M1, the determination of the partial $PSF$ was not thoroughly addressed. When
an image contains sufficiently isolated point sources, their shape can be used
to determine an accurate $PSF$. However, this simple $PSF$ determination is rarely
possible in crowded fields, which generally contain no star sufficiently
isolated for this purpose.

Magain et al.  (2006, hereafter M2) have thus developed a version of the algorithm allowing to
\emph{simultaneously} perform a deconvolution and determine an accurate $PSF$ in
fields containing exclusively point sources, even if no isolated star can be found. 
It relies on the minimization of the following merit function: 
\begin{equation}
\label{eq:aab}
S = \sum_{i=1}^N \frac{1}{\sigma_i^2} (d_i - [s \ast f]_i)^2 + \lambda H(s)
\end{equation}
  where $\ast$ stands for the convolution operator, $N$ is the number of pixels within the image, $d_i$ and $\sigma_i$ are the measured intensity and standard deviation in pixel $i$, $s_i$ is the {\em unknown} value of the partial $PSF$ and $f_i$ is the intensity of the deconvolved image in pixel $i$. $H(s)$ is a smoothing constraint on the $PSF$ which is introduced to regularize the solution and $\lambda$ is a Lagrange parameter. This algorithm performs an \emph{optimal} $PSF$ determination, in the sense that it uses for  this purpose the whole information available in the image. It relies on the assumption that the $PSF$ is constant over the image.  To extend the validity of this assumption, one can treat relatively small sub-images if $PSF$ variations are suspected. The $PSF$ determination is decomposed in several steps and is optimized in order to avoid including faint blending stars in the $PSF$ wings, allowing their detection after inspection of the deconvolved image and the residuals map (see M2 for more details). Taking into account the blending stars which are undetectable in the original image results in a better accuracy on the $PSF$, and thus on the astrometry and photometry.

The  deconvolved light distribution $f$ may be written:
\begin{equation}\label{eq:aac}
f(\boldsymbol{x}) = \sum_{k=1}^M a_k r(\boldsymbol{x} - \boldsymbol{c}_k)
\end{equation} where $M$ is the number of points sources in the image,
$r(\boldsymbol{x})$ is the final $PSF$ (fixed) while $a_k$ and $\boldsymbol{c}_k$ are free parameters corresponding to the intensity and position of point source number $k$. Note that the right-hand side of (\ref{eq:aac}) represents only point sources, thus the sky background is supposed to be
removed beforehand, and it is assumed that the data do not contain any extended source.

\subsection{Optimization of the algorithm for transit photometry}
\paragraph{Increase of the processing speed}

If we consider an image with $N$ pixels, containing $M$ point sources, we are
left with the problem of determining $N + 3 M$ parameters, i.e. $N$ pixel
values of the partial $PSF$ and 3 parameters for each point source (one
intensity and two coordinates). In follow-up transit photometry, we have
generally to analyze several hundreds of images for a single transit.
Furthermore, we are not allowed to analyze only a small fraction of the image
around the target star. Indeed, several systematic noise sources exist, mainly
due to atmospheric effects, and the correction of the light curve of the
analyzed star by the mean light curve of several comparison stars is needed to
tend towards a photon noise limited photometry. This implies that the
 $S/N$ (signal-to-noise ratio) of the comparison light curve must be
significantly higher than the $S/N$ of the target star.  Thus, we have to
analyze a field of view large enough to contain many reference stars (but yet
smaller  that the coherence surface of the systematics). In
practice, we thus need to process several hundreds of images containing dozens
or even hundreds of point sources each. As the deconvolution of such an image
with the algorithm presented in M2 can last up to one day for a very crowded field 
with an up-to-date personal computer, we have to increase  drastically  the
processing speed.

To reach this goal, we use three bits of {\em prior knowledge}.\\
-- First, we know that our hundreds of images correspond to the same field of
view and, thus, contain the same objects.\\
-- Secondly, we can assume that the stellar positions do not change during the
observing run.  As we are not interested in the astrometry of the stars but 
only in their photometry, we can use the best seeing image or a combination of
the best-quality images as a reference frame and analyze it with the standard
algorithm in order to obtain the astrometry, which is kept fixed during
the rest of the analysis.  All we still need to know to obtain the positions of the
stars in each image is the amount by which this image is translated with
respect to the reference image.  This translation is simply determined by a
cross--correlation of the images. We neglect image stretch as in practice we treat 
several relatively small sub-images to improve the validity of a constant $PSF$ assumption.
The result is that, for each point source, we are left with only one free parameter 
(its intensity) instead of three.\\
-- The third {\em prior knowledge} is that the relative intensities of most
point sources do not change much from one image to another. We can thus obtain
a first approximation of the partial $PSF$ by assuming that the relative
intensities of the point sources are identical to those in the reference
image, just allowing for a
common scaling factor on the whole image, which takes into account variations 
of atmospheric transparency, airmass, exposure time, etc.
 We then use this first partial $PSF$ estimate to obtain a better approximation
of the intensities.  With these better intensities, we redetermine an improved
partial $PSF$, and so on until convergence.

The most time-consuming task in the standard  algorithm is the iterative
determination of the point sources' positions and intensities.  Here, we
already save a large amount of computing time by keeping the positions fixed. 
Moreover, when the only unknowns are the point sources' intensities, the
problem becomes {\em linear} in all the parameters.  We thus have to solve a
set of $M$ linear equations, which can be done directly, without any
iteration. However, as the direct solution of this set of equations is quite
unstable, we use the Singular Value Decomposition method (SVD, \cite{Press1}),
which has been found to give excellent results.

The analysis of a set of images is thus divided into two parts. In the first
one, a reference image is deconvolved in order to obtain the astrometry and
starting values for the point source intensities. Then, the shift of each image
in the set relative to the reference frame is determined by
cross--correlation.  In the second part, an initial $PSF$ is determined for each
image, using the fixed astrometry (including shift) and photometry.  Improved
source intensities are then obtained by solving the linear problem.  As the
accuracy of the partial $PSF$ depends on the accuracy of the photometry and
vice-versa, the process is
repeated several times until convergence.  In practice, the convergence is
reached after a maximum of 5 cycles.

\paragraph{Determination of the sky background}

A tricky problem in crowded field photometry is the determination of the
sky background.  Fitting a rather smooth surface through seemingly ``empty"
areas may lead to seeing-dependent systematic errors.  A much more robust
method consists in determining the sky background level so that the shape of
all point sources remains the same, irrespective of their intensities and
positions. 
Indeed, a wrong sky level would affect weaker sources much more strongly than
brighter ones. The fact that our method forces all point sources to have the
same $PSF$ shape can thus be used to obtain an accurate determination of the sky
background. 

In practice, this is very simply done by not subtracting the sky background
prior to
processing, but rather by implementing its determination into the method. In
this
case, the observed light distribution $d$ can be modelled as:
\begin{equation}\label{eq:aae}
d(\boldsymbol{x}) = s(\boldsymbol{x})\ast \sum_{k=1}^M a_k r(\boldsymbol{x} -
\boldsymbol{c}_k) + b(\boldsymbol{x})
\end{equation} where the  sky background is represented by the function
$b(\boldsymbol{x})$, chosen to be relatively smooth. A 2-dimension second
order polynomial (6 free parameters) was found suitable for images
obtained in the optical.

For the deconvolution of the reference image, we have now to minimize the
following merit function: \begin{equation}\label{eq:aaf}
S = \sum_{i=1}^N \frac{1}{\sigma_i^2} (d_i - b_i-[s \ast f]_i)^2 + \lambda H(s)
\end{equation} 
where $b_i$ is the sky level in pixel $i$.

For the deconvolution of the complete set of images, the coefficients of $b(\boldsymbol{x})$
are determined by adding an extra step to the analysis of the whole set of
images, in each iteration.  Using the previous approximation of the partial
$PSF$ and point source
intensities, we determine the polynomial coefficients of the sky background by
simple SVD solution of a linear set of equations where all parameters are
fixed but the coefficients of $b(\boldsymbol{x})$. The whole process (1: sky background, 2:
partial $PSF$, 3: intensities determination) is repeated until convergence.

\section{Results} 
\subsection{Light curve analysis}

The light curves obtained with our deconvolution-based photometry algorithm are shown
in Fig. 1. The flux variations before the second
transit are intruiging, but we remarked that they are correlated to
the location of a bad column of the CCD close to OGLE-TR-113 and
a bright reference star. In fact, the bad column is located on their $PSF$s in
the first 101 images, and it moved away drastically a few exposures before the
transit,  so the rest of the light curve is reliable. The 101 first points
were not used in the transit fitting.

For the first night, the dispersion of the light curve of OGLE-TR-113 before
the transit is 1.20 mmag, while the mean photon noise is 0.95 mmag. For the
second night, the dispersion of the light curve after the transit is 1.26
mmag, for the same mean photon noise. The slightly higher dispersion for the second
night can be explained by the increased seeing and the fact that OGLE-TR-113 has a 0.4 mag brighter
visual companion about 3$\arcsec$ to the South (see Fig. 2). When a star's $PSF$ is
blended with another one, a part of the noise of the contaminating star
is added to its own noise, resulting in a decrease of the maximal
photometric accuracy attainable. This effect is of course very dependent on the
seeing, and may have a large impact on the final harvest of a transit survey
(\cite{Gillon1}). As the average seeing was higher during the second
night, we thus expect a lower accuracy for this sequence.
Nevertheless, the obtained accuracies for both nights can be judged as
excellent. 

Our method has the advantage to produce higher resolution images which can be
used to detect a faint blending companion around a star which could not be seen
on a lower resolution image. As shown in Fig. 2, there is no evidence of such faint
companions around OGLE-TR-113 in our results.

\begin{figure}
\label{fig:a}
\centering                     
\includegraphics[width=9.0cm]{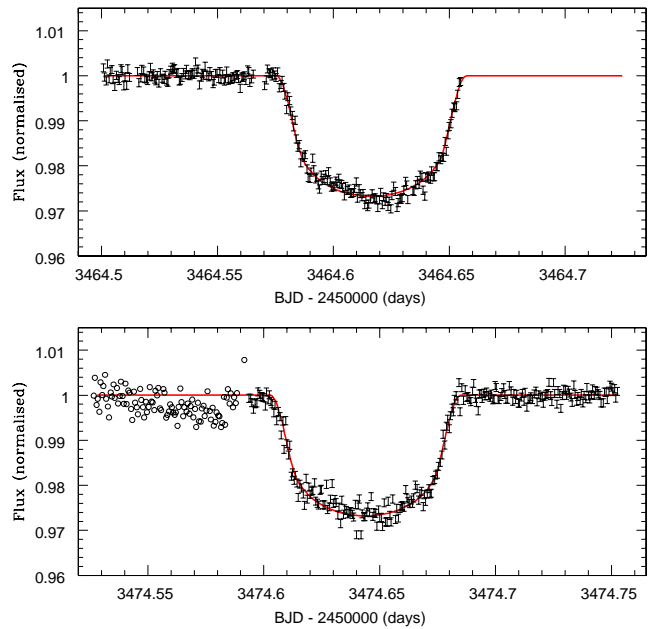}
\caption{Our NTT/SUSI2 light curve for the first ($top$) and the second ($bottom$) observed transits of
OGLE-TR-113b, with the best fit transit curve superimposed. For the second transit, the variations of
the flux before the transit are due to a bad column of the CCD
located close to the $PSF$ cores of OGLE-TR-113 and a bright reference star (open
symbols).}
\end{figure}

\begin{figure*}
\label{fig:bbc}
\centering                     
\includegraphics[width=18.0cm]{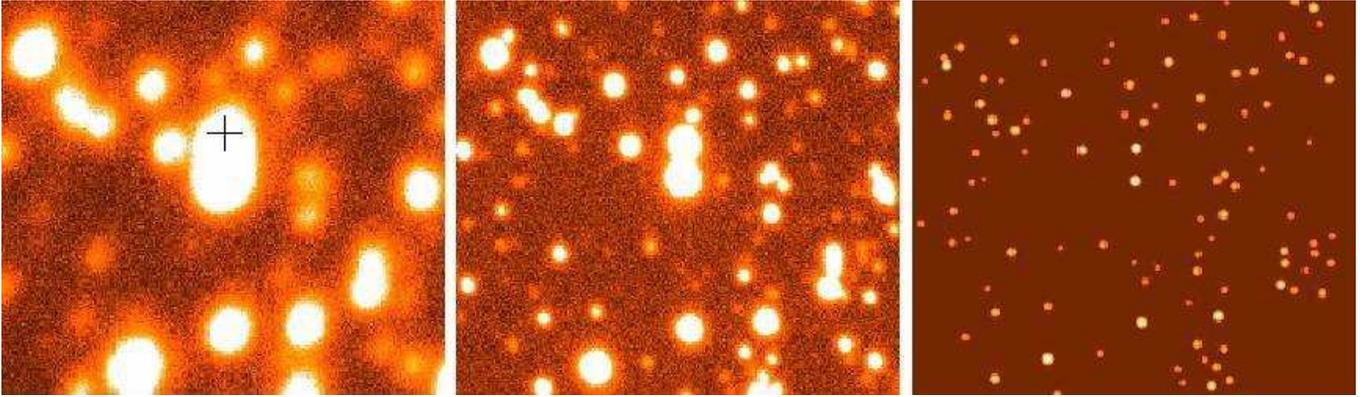}
\caption{OGLE-TR-113 (marked with a cross) in a 256 pixels $\times$ 256 pixels sub-image (0.7 $\arcmin$ $\times$ 0.7 $\arcmin$) from
 the worst (\emph{left}) and best (\emph{middle}) seeing NTT/SUSI2 image of the run ($top$ = North, $left$ = East).
The nearby star just South of OGLE-TR-113 is about 0.4 mag brighter. \emph{Right}: deconvolved image. At
this resolution (2 pixels = 0.32$\arcsec$), no trace of other faint companions is visible.}
\end{figure*}
\subsection{Transit fitting}
The transit fitting  was performed with  transit curves computed with the
procedure of Mandel \& Agol (\cite{Mandel1}), using quadratic limb-darkening
coefficients. The transit parameters were obtained in two iterations. A
preliminary solution was first fitted to determine epochs for  NTT photometric
series. A period was then determined by comparing these epochs with OGLE-III
ephemeris (\cite{Konacki2}), allowing a very high accuracy  on the period due
to the large time interval separating the two sets of measurements (795 times
the orbital period). The period was then fixed to this value and the radius
ratio, orbital inclination, transit duration and transit epoch were fitted by
least squares using the NTT data. The limb-darkening coefficients used were 
$u_1$ = 0.55 and $u_2$ = 0.18, obtained from Claret (2000) for the following stellar
parameters: effective temperature $T_{eff}$ = 4750 K, metallicity $[M/H]$ = 0.1, surface gravity log $g$ = 4.5 and microturbulence velocity $\xi_t$ = 1.0
km/s, based on the parameters presented in Santos et al. (2006).

To obtain realistic uncertainties for the fitted transit parameters, it is
essential to take into account the correlated noise present in the light
curves, as shown by Pont et al. (\cite{Pont2}). Although we have attained a
very good level of stability in our photometry, the residuals are not entirely
free of covariance at the sub-millimag level. We model the covariance of the
noise from the residuals of the light curve itself. We estimate the amplitude
of systematic trends in the photometry from the standard deviation over one
residual point, $\sigma_1$ , and from the standard deviation of the sliding
average of the residuals over 10 successive points,  $\sigma_{10}$. The
amplitude of the white noise  $\sigma_w$ and the red noise $\sigma_r$ can then
be obtained by resolution of the following system of 2 equations:
\begin{eqnarray}\label{eq:zza}
\sigma_1^2 & = & \sigma_{w}^2 + \sigma_{r}^2\\
\sigma_{10}^2 & = & \frac{\sigma_{w}^2}{10} + \sigma_{r}^2
\end{eqnarray} We obtained $\sigma_r$ = 400 $\mu$mag for both nights. We assume that a systematic feature of
this amplitude could be present in the data over a length similar to the transit duration. Therefore,  if $\chi_{bf}$ is the $\chi^2$ of the best fit,
instead of using $\Delta \chi^2$ = 1 to define the 1-sigma uncertainty
interval,
we use 
\begin{equation}\label{eq:zzb}
\Delta \chi^2 = 1 + N_{tr,i}\frac{\sigma_{r,i}^2}{\sigma_{w,i}^2}
\end{equation} 
for each individual transit, where $N_{tr,i}$ is the number of points in the
transit $i$. As the
residuals between different transits are not correlated, combining the data
from both individual transits gives:
\begin{equation}\label{eq:aag}
\Delta\chi^2  = \sqrt{\Delta\chi_1^2 + \Delta\chi_2^2}
\end{equation} 

Exploring the parameters space for $\chi^2 =$ $\chi^2_{bf}+ \Delta\chi^2$, we
then estimated the uncertainties on our parameters. 

The results are given in Table \ref{tab:caa}. The fit of the final light
curve on the NTT data is shown in Fig. 3. This figure also shows the final transit curve superimposed on the OGLE-III data (the limb-darkening coefficients are
changed, as OGLE-III observations have been obtained in the $I$ filter), and  
the phased NTT/SUSI2 data after binning on 2 points superimposed on the  best-fit transit curve. For this binned light curve, 
the dispersion before the transit is $\sim$800 $\mu$mag (first night), and $\sim$850 $\mu$mag after the transit (second night). 

\begin{figure*}
\label{fig:bbf}
\centering                     
\includegraphics[width=18.0cm]{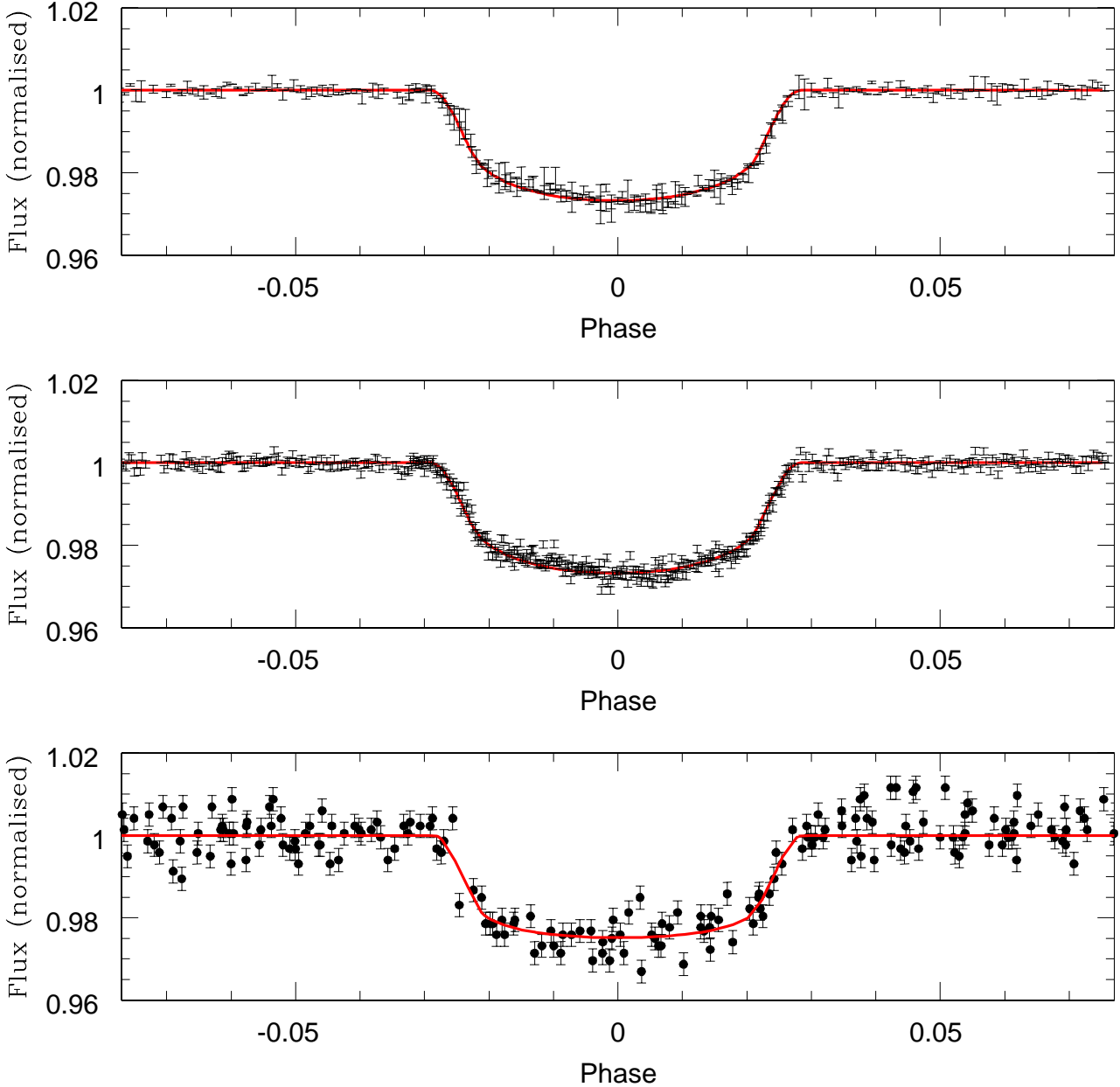}
\caption{The best-fit transit curve is shown together with the phased NTT/SUSI2
data after binning on 2 points (\emph{top}), without binning (\emph{middle}) and 
 with the phased OGLE-III data (\emph{bottom}). }
\end{figure*}

\subsection{Radius and mass determination}
Combining the constraints from our new transit curves (dependence on $R_{\ast}
M_{\ast}^{-1/3}$) and from the spectroscopic determination of $T_{eff}$, log
$g$ and [Fe/H] (\cite{Santos1}), we computed the radius and mass of
OGLE-TR-113 as in  Bouchy et al. (2005), taking the relation between $M_{\ast}$,
$R_{\ast}$, and the atmospheric parameters from an interpolation of Girardi et
al. (\cite{Girardi1}) stellar evolution models. 
The value of the planetary radius was then derived from the radius ratio and
from $R_{\ast}$. Next, we fitted a sinusoidal orbit by least--squares to the
radial velocity data with the new period and epoch, obtaining the planetary
mass from $M_{\ast}$ and the semi-amplitude of the radial velocity orbit. 

Our values for the radius and mass of OGLE-TR-113 and its planetary companion
are given in Table \ref{tab:caa}, which also presents the  values obtained by 
Bouchy et al. (2004) and Konacki et al. (2004). Our results are in good agreement
with the previous studies, but the uncertainties on the mass and  on the 
radius of the planet are significantly lower.  In fact, our high photometric
accuracy allows to reach 
the regime where the uncertainties on the mass and radius of the primary dominate: 
the use of other stellar evolution models should introduce parameter changes
which are of the same order than the error bars. The development of this point is 
beyond the scope of this paper.
\begin{table*}
\centering
\begin{tabular}{cccc}
\hline
 & {\bf A} & B  & C \\
\hline 
Inclination angle[deg] & {\bf $88.8 - 90$} & 85 - 90 &  $88.4 \pm 2.2$\\
Period[days] & {\bf  $1.4324757 \pm 0.0000013$} & $1.43250$ (adopted) & $1.4324758 \pm 0.0000046$  \\
Semi-major axis[AU] & {\bf  $0.0229  \pm 0.0002$}  & $0.0228  \pm 0.0006$ &   $0.02299  \pm 0.00058$\\
Eccentricity  (fixed) &{\bf $0$} & 0 & 0  \\
& \\
Primary mass[$M_{\odot}$] & {\bf  $0.78 \pm 0.02$} & $0.77 \pm 0.06$ & $0.79 \pm 0.06$ (adopted)\\
Primary radius[$R_{\odot}$] &{\bf   $0.77 \pm 0.02$} & $0.765 \pm 0.025$ &  $0.78 \pm 0.06$ (adopted)\\
& \\
Planet mass[$M_J$] &{\bf   $1.32 \pm 0.19$} & $1.35 \pm 0.22$ & $1.08 \pm 0.28$ \\
Planet radius[$R_J$] & {\bf  $1.09 \pm 0.03$} & $1.08^{+0.07}_{-0.05}$ & $1.09 \pm 0.10 $ \\
Planet density[g cm$^{-3}$] &{\bf   $1.3 \pm 0.3 $} &  $1.3 \pm 0.3 $ & $1.0 \pm 0.4 $ \\
\hline
\end{tabular}
\caption{Parameters obtained from this analysis ({\bf A}) for OGLE-TR-113 and its
planetary companion, compared to the ones presented in Bouchy et al. (2004) (B) and in Konacki et al. (2004) (C). }
\label{tab:caa}
\end{table*}

\begin{table}
\centering
\begin{tabular}{cc}
\hline
$T_{0,OGLE}$[BJD]& $2 452 325.79823 \pm  0.00082$\\
$T_{0,1}$[BJD] & {\bf $2 453 464.61665 \pm  0.00010$} \\
$T_{0,2}$[BJD] & {\bf  $2 453 474.64348 \pm  0.00017$} \\
\hline
\end{tabular}
\caption{Ephemeris of OGLE-TR-113b transits. $T_{0,OGLE}$ is from Konacki et al. (2004), while $T_{0,1}$ and $T_{0,2}$ are for the two transits analyzed in this work.}
\label{tab:cab}
\end{table}

\subsection{Transit timing}
OGLE-III (\cite{Konacki2}) and our new NTT transits ephemeris are presented in Table \ref{tab:cab}. 
Using these ephemeris and
assuming the absence of long-term transit period variations due, e.g., to the
oblateness of the star or dissipative tidal interactions between the planet and
the star, we obtained a value for the period in perfect agreement with the one presented in Konacki et al. (2004), but 
with a much smaller error bar.
Our extremely high accuracy on the orbital period ($\sim$
0.1 s) is mainly due to the long delay between OGLE-III and NTT observations. Moreover, our accuracy on the epoch 
of the NTT transits is $\sim$ 12 s, while the accuracy on the OGLE-III epoch is $\sim$ 71 s. 

Short-term transit timing
variations ($TTV$) may be induced by the presence of a satellite  (\cite{Schneider1})
or a second planet (\cite{Miralda1}, \cite{Agol1}, \cite{Holman1}). 
If we examine the
temporal deviation of the second NTT  transit off the expectation based on our
period and the first NTT transit time, we obtain a value of $\sim 43$ s. This $TTV$ has 
a  statistical significance of 2.5 sigmas. 
Nevertheless, we must stay cautious 
because systematics are able to slightly distort 
light curves, all the more so since these transit light curves are not complete. 
This is the case here: we lack the flat part after the first transit and, 
due to the bad column, the flat part prior to the second one. A way to estimate the
likelihood of the hypothesis that systematics are responsible for the
observed $TTV$ is to determine the transit ephemeris for the 
ingress and egress independently. Table \ref{tab:cac} and Fig. 4 show the result. For both transits, transit times 
obtained from ingress and egress are in good agreement, leading us to reject such systematic errors as the explanation for the observed $TTV$. 

\begin{table}
\centering
\begin{tabular}{ccc}
\hline
$T_{0,1,ingress}$[BJD] &  2453464.61669 $\pm$ 0.00014 \\
$T_{0,1,egress}$[BJD] & 2453464.61653 $\pm$ 0.00014 \\
$T_{0,2,ingress}$[BJD] &  2453474.64299 $\pm$ 0.00024 \\
$T_{0,2,egress}$[BJD] & 2453474.64400 $\pm$ 0.00024 \\
\hline
\end{tabular}
\caption{Ephemeris obtained from ingress and egress of both transits.}
\label{tab:cac}
\end{table}

\begin{figure}
\label{fig:bbp}
\centering                     
\includegraphics[width=8.0cm]{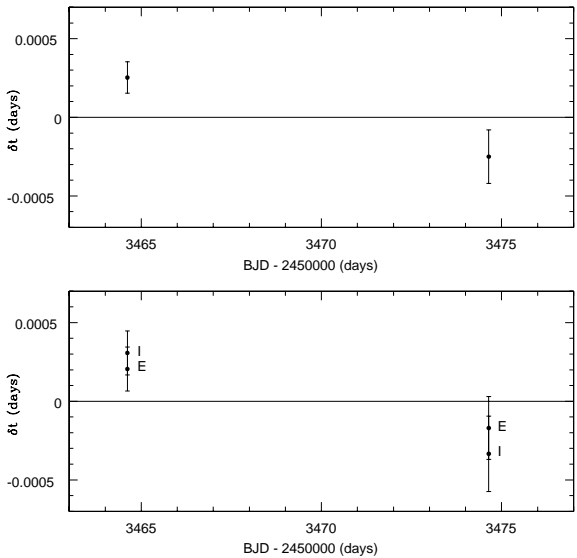}
\caption{This figure shows the agreement between the transit times obtained from the complete light curves (\emph{top}) and using ingress (I) and egress (E) independently (\emph{bottom}) and the predicted transit times based on the determined period and OGLE-III ephemeris.}
\end{figure}

The statistical significance of the $TTV$ is rather low, as can be seen in Fig. 4. 
Future observations are needed to confirm its existence.  
We notice nevertheless that our timing accuracy would clearly be good enough to allow the 
detection of a third planet or a satellite giving rise to a $TTV$ amplitude of
 1 minute. Considering the case of OGLE-TR-113b and, as the cause of such a $TTV$, a satellite  with an orbital distance to the planet equal to the Hill radius, we can obtain an estimate of its mass by using the formula
(\cite{Schneider1}):
\begin{equation}\label{eq:aah}
M_s \sim \pi \frac{\delta t}{P_p} \biggl(\frac{3 M_{\ast}}{M_p}\biggr)^{1/3}
M_p
\end{equation} where $M_{\ast}$, $M_p$ and $M_s$ are respectively the masses of
the star, planet and satellite, $P_p$ is the orbital period of the
planet and $\delta t$ the amplitude of the $TTV$. We obtain   $M_s \sim 7 M_{\oplus}$.  

For an exterior pertubing planet, the most interesting case would be a pertubing planet in 2:1
mean--motion resonance with OGLE-TR-113b, for which we obtain with the
following formula (\cite{Agol1}): 
\begin{equation}\label{eq:aaj}
M_{p2} = 4.5 \frac{\delta t}{P_{p1}} M_{p1}
\end{equation} a mass $\sim  1  M_{\oplus}$.

These computations demonstrate the interest of high accuracy photometric 
follow-up of known transiting exoplanets and show that the  accuracy obtained
with our reduction method for high quality data would allow
the detection of very low mass objects.

\section{Conclusions}
The results presented here show that our  new photometry algorithm 
is well suited for follow-up transit photometry, even in very crowded fields. After analysis  of NTT
SUSI2 observations of two OGLE-TR-113b transits, we have obtained two very high
accuracy transit light curves with a low level of systematic residuals. Combining our new photometric
data with OGLE-III ephemeris, spectroscopic data and radial velocity measurements, we have determined planetary 
and stellar parameters in excellent agreement with
the ones presented in Bouchy et al. (2004) and Konacki et al. (2004), but significantly more precise.
We notice that the sampling in time, the sub-millimag photometric accuracy and the systematics residuals level of our light curves 
would be good enough to allow the photometric detection of a 
transiting Hot Neptune,  in the case of a small star as OGLE-TR-113.

We have obtained a very precise determination of the transit times, and, combining
them with OGLE-III ephemeris, we could determine the orbital period with a very high
accuracy. The precisions on the epochs and the period would in fact be high enough
to allow the detection of a second planet
or a satellite, for some ranges of orbital parameters and masses. Even a
terrestrial planet could be detected with such a transit timing precision.

\begin{acknowledgements} 
The authors thank the ESO staff on the NTT telescope at La Silla for their diligent and competent help during the observations. MG acknowledges support by the Belgian Science Policy (BELSPO) in the framework of the PRODEX Experiment Agreement C-90197. FC acknowledges financial support by the Swiss National Science Foundation (SNSF).
\end{acknowledgements}

\end{document}